# RLC series circuit made simple and portable with smartphones

*Ives Torriente-García[1], Arturo C. Martí[2], Martín Monteiro[3], Cecilia Stari[2], Juan C. Castro-Palacio[1], Juan A. Monsoriu[1]*

[1]Centro de Tecnologías Físicas, Universitat Politècnica de València, 46022 València, Spain

[2]Instituto de Física, Universidad de la República, 11400 Montevideo, Uruguay

[3]Universidad ORT Uruguay, 11100 Montevideo, Uruguay

Abstract

This article presents a novel method for studying RLC series circuits using two smartphones, one used as a signal generator and the other as an oscilloscope. We measure the voltage at the external resistor as a function of frequency when subjected to a sinusoidal electromotive force. The experimental results demonstrate a remarkable agreement with the theoretical curve for the voltage at the resistor and the resonance frequency, validating the accuracy of the smartphone-based setup. The experiment fills the gap in educational materials related to electrical circuits and provides a portable alternative to traditional, expensive laboratory equipment.

**Introduction**

Physics experiments using smartphone sensors are very common to see nowadays for both secondary and introductory Physics courses at the university level.[1] These experiments are proven to foster the motivation and independence of students what positively impacts their academic performance. However, in this broad range of works some topics are more represented than others, for instance, mechanics,[2,3] oscillations,[4] beats,[5] acoustics,[6] and optics.[7] The published work on electromagnetism using smartphone sensors involves mostly the measurement of magnetic fields.[8-10] However, experiments relating electrical circuits appear in a much lesser extent.[11,12]

Direct and alternating current circuits combining resistors, capacitors and coils such as RC, RL, and RLC series circuits are commonly seen in secondary and introductory physics courses.[12-20] Most published work on simple circuits includes the study of charge and discharge of a capacitor in an RC series circuit,[12,14] and current variations upon connecting and disconnecting an RL series circuit.[15,16] With regard to RLC circuits,[17-20] we would like to point out a very simple work[20] where the resonance curve (voltage at the capacitor versus the frequency) was obtained using a voltmeter instead of an oscilloscope. We go beyond in this work and substitute the signal generator and the oscilloscope by smartphones. This idea was first proposed in reference.[12] In this article, authors used the headphone port designed for connecting an external microphone and speakers to read data from an external circuit. On the other hand, in reference[21] authors proposed to transform a smartphone in an arbitrary signal generator by means of using a java code for Android. Here we make it simple by using the *Tone Generator* option available in the free app for Android Physics Sensors Toolbox Suite.[22] Another way of turning a smartphone into an oscilloscope takes the advantage of Arduino boards to communicate with a smartphone using the bluetooth interface,[23] which is a valuable option for smartphones not including a headphone port.

In this paper we present a simple method to study an RLC series circuit connected to a sinusoidal electromotive force. To achieve simplicity and portability, we utilized two smartphones, one as a signal generator and the other as an oscilloscope, replacing the expensive equipment commonly found in physics laboratories. By employing smartphones, we have not only reduced the cost of the experiment but also made it accessible outside the confines of the physics laboratory, providing students with the opportunity to conduct the study in various settings.

**The experiment**

Figure 1 shows the experimental setup used in this study. A picture of the actual setup is shown on the left-hand side, and a schematic representation of the circuit is shown on the right-hand side. A Samsung Galaxy S52 smartphone was used as the signal generator and another, a smartphone Redmi Note 7, is used as the oscilloscope. The RLC circuit is formed by a resistor of resistance $R = (220.0 \pm 4.4)\,\Omega$ (manufacturer value), an inductor or coil of inductance $L = (10.37 \pm 0.25)$ mH (measured with a digital LRC meter, model PROSTER BM4070) and a capacitor of capacitance $C = (1.00 \pm 0.01)\,\mu F$ (manufacturer value). The internal resistance of the smartphone used as signal generator was $R_i = (10.620 \pm 0.002)\,\Omega$ when measured with a fixed multimeter Keysight 34461A. The ohmic resistance of the coil, $R_L = (19.381 \pm 0.003)\,\Omega$, was also measured with the same equipment. $R_T = R + R_i + R_L = (250.0 \pm 4.4)\,\Omega$ is the total ohmic resistance of the circuit. The maximum voltage at the resistor which is directly proportional to the maximum current (see Eq. 3 below) was measured with another smartphone used as oscilloscope.

The signal generation capability is achieved using the *Tone Generator* included in the free Android app Physics Sensors Toolbox Suite.[22] The *Tone Generator* can be found on the left panel of the main screen of the app. The frequency (0 - 20000 Hz) and type of signal (sinusoidal, square, saw-tooth, triangular) are the parameters available to set up. The amplitude of the generated wave can be controlled by changing the volume of the phone. An audio jack has been used to allow the signal out. On the other hand, the second smartphone was used as oscilloscope by means of the option *Audio Scope* included in the free Android app phyphox.[24] This option visualizes any audio sent in using the jack connector, which means in this paper the voltage at the external resistor. We should point out that not all smartphones perform well as oscilloscopes therefore we recommend to first check that the measured signal is not deformed. The free app(s) used in this work can be seen on the screenshots shown in Fig. 1. The picture was taken while the app(s) were being used.

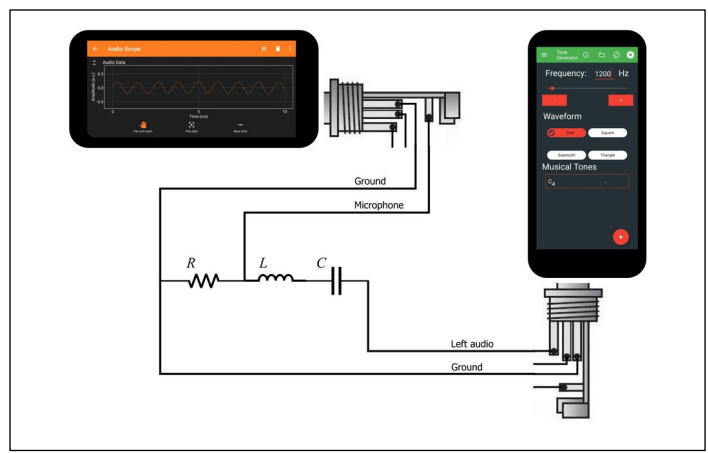

**Fig. 1.** Schematic representation of the experimental setup. The free app(s) used in this work can be seen on the screens of the smartphones.

## Results

First, we have studied the range of voltages at the audio port when a sinusoidal wave is generated with Physics Sensors Toolbox Suite[23]. For this purpose, we have used a conventional oscilloscope (model Tektronix TBS 1072B-EDU) to measure the amplitude of the electromotive force ($\varepsilon_{max}$) for each step of the volume button of the smartphone. Figure 2 shows the values of $\varepsilon_{max}$ as a function of the number of steps of the volume button (a total of n = 15 steps). The measures of $\varepsilon_{max}$ have been taken for a low (250 Hz) and high value (12 kHz) of frequency within the range used in this work (0 – 12 kHz) and results are largely the same (Figure 2). For all measurements, the volume has been set at step n = 7 (out of 15 steps available) which corresponds to 32.4 mV. Keeping the volume step at n = 7, we also measured the voltage with phyphox app to obtain 0.516 a.u. (arbitrary units). It means that the a.u. to volt conversion is 0.062 V/a.u.

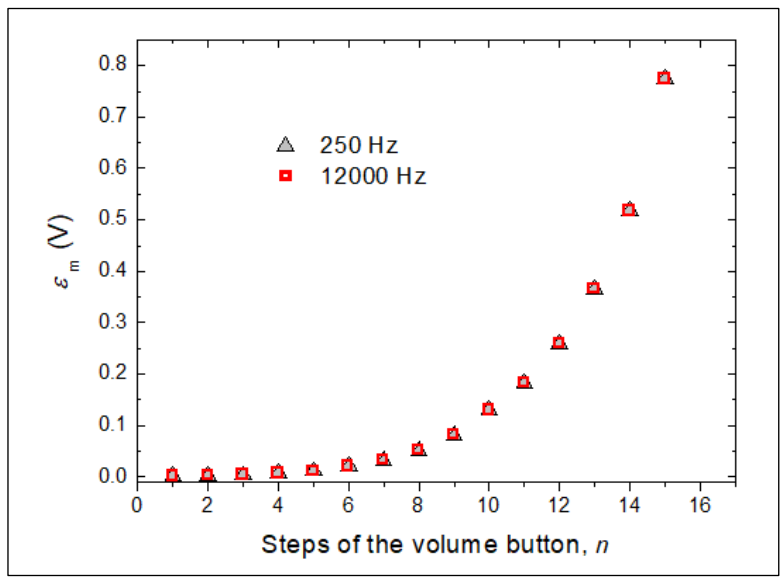

**Fig. 2.** Maximum electromotive force at the audio port of the smartphone ($\varepsilon_{max}$) as a function of the number of steps of the volume button. The values have been measured using a conventional oscilloscope (model Tektronix TBS 1072B-EDU)

An important phenomenon when it comes to the study of RLC series circuits is resonance. It is given when the capacitive reactance ($X_C$) and the inductive reactance ($X_L$) are equal. Then, the general impedance becomes a real number which is equal to the ohmic resistance of the circuit $Z = R_T$. For the resonance frequency the maximum current is maximal, $I_m = \varepsilon_m / R_T$. We can know about the behaviour of the current by looking at the voltage at the resistor as these to quantities are directly proportional (Eq. 3).

Table 1 registers the values of the maximum voltage at the resistor ($V_{Rm}$) measured with the smartphone used as oscilloscope as a function of the frequency ($f$) of a sinusoidal wave used as input and originated at the smartphone used as signal generator. These values are shown in Fig. 2.

**Table I.** Values of the maximum voltage at the resistor ($V_{Rm}$) as a function of the frequency ($f$) of the sinusoidal wave used as input in the RLC series circuit.

| $f$ (Hz) | $V_{Rm}$ (a.u.) | $f$ (Hz) | $V_{Rm}$ (a.u.) | $f$ (Hz) | $V_{Rm}$ (a.u.) |
|---|---|---|---|---|---|
| 100 | 0.0540 | 1650 | 0.4387 | 5000 | 0.2726 |
| 200 | 0.1260 | 1700 | 0.4372 | 5500 | 0.2559 |
| 300 | 0.1820 | 1750 | 0.4374 | 6000 | 0.2370 |
| 400 | 0.2323 | 1800 | 0.4395 | 6500 | 0.2207 |
| 500 | 0.2750 | 2000 | 0.4317 | 7000 | 0.2030 |
| 600 | 0.3162 | 2200 | 0.4199 | 7500 | 0.1968 |
| 700 | 0.344 | 2400 | 0.4072 | 8000 | 0.1864 |
| 800 | 0.3676 | 2600 | 0.3958 | 8500 | 0.1785 |
| 900 | 0.3845 | 2800 | 0.3875 | 9000 | 0.1706 |
| 1000 | 0.4048 | 3000 | 0.3796 | 9500 | 0.1595 |
| 1100 | 0.4182 | 3200 | 0.3612 | 10000 | 0.1550 |
| 1200 | 0.4243 | 3400 | 0.3532 | 10500 | 0.1470 |
| 1300 | 0.4315 | 3600 | 0.3416 | 11000 | 0.1324 |
| 1400 | 0.4378 | 3800 | 0.3307 | 11500 | 0.1294 |
| 1500 | 0.4405 | 4000 | 0.3155 | 12000 | 0.1246 |
| 1600 | 0.4410 | 4500 | 0.2958 | | |

The maximum current established at the RLC series circuit when it is powered with a sinusoidal electromotive force ($\varepsilon$) is calculated as:

$$I_m = \frac{\varepsilon_m}{|Z|}, \tag{1}$$

where $\varepsilon_m$ is the maximum electromotive force and $|Z|$ is the modulus of the complex impedance. By substituting the expression for the modulus of the impedance the above equation becomes:

$$I_m = \frac{\varepsilon_m}{\sqrt{R_T^2 + (X_L - X_C)^2}} = \frac{\varepsilon_m}{\sqrt{R_T^2 + (2\pi f L - 1/(2\pi f C))^2}}, \tag{2}$$

Where $L$ is the inductance of the coil, $C$ the capacitance of the capacitor, and $R_T = R + R_i + R_L$ is the total ohmic resistance. As mentioned above, the terms $R$, $R_i$ and $R_L$ represent the resistance of the external resistor, the internal resistance of the smartphone used as generator and the ohmic resistance of the inductor, respectively. The maximum voltage at the resistor which depends on the frequency can be represented as:

$$V_{Rm}(f) = I_m R = \frac{\varepsilon_m R}{\sqrt{R_T^2 + (2\pi f L - 1/(2\pi f C))^2}} \tag{3}$$

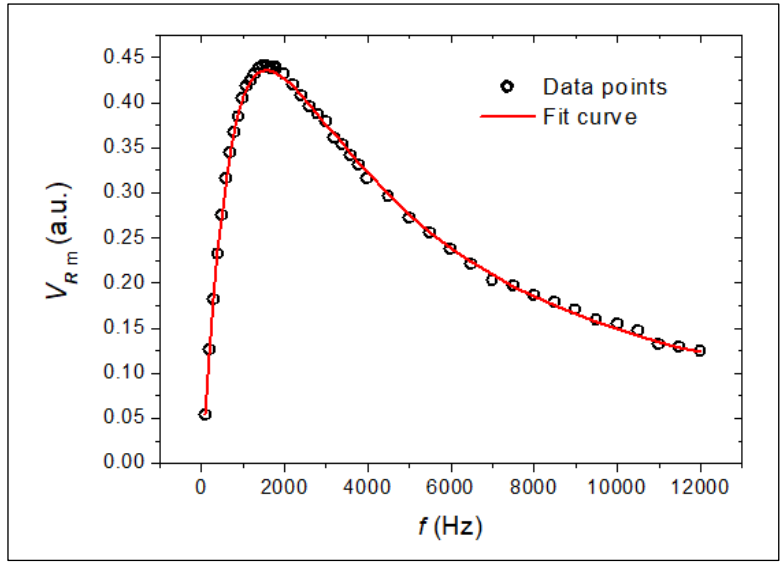

**Fig. 3.** Maximum voltage at the resistor ($V_{Rm}$) as a function of the frequency ($f$). The data points (open circles) are shown along with the fitting curve using the theoretical expression (red solid line).

In order to test the suitability of the experimental setup to study the RLC series circuit, the expression of $V_{Rm}(f)$ in Eq. (3) has been fitted to the experimental data (red solid line in Figure 2). Two parameters have been left free, namely, the maximum electromotive force ($\varepsilon_m$) and the resonance frequency ($f_0$). The resulting fitted value for the maximum is $V_{Rm} = (0.452 \pm 0.003)$ a.u., and for the resonance frequency $f_0 = (1560.5 \pm 5.9)$ Hz. The overall quality of the fitting is good with a correlation coefficient of 0.999. By using the fitted resonance frequency and the value reported by the manufacturer for the capacitance, the inductance of the coil can be calculated from eq. 3 as $L = (1/C)\left[1/(2\pi f_0)\right]^2 = (10.4 \pm 0.1)$ mH which compares very well to the previously measured value $L = (10.37 \pm 0.25)$ mH. On the other hand, using the fitted maximum electromotive force and the value reported by the manufacturer for the resistance, the total ohmic resistance of the circuit can be calculated from eq. 3 at resonance, $R_T = R \dfrac{\varepsilon_m}{V_{Rm}} = (251 \pm 8)\,\Omega$, which compares very well to the total ohmic resistance of the circuit previously mentioned, $R_T = (250.0 \pm 4.4)\,\Omega$.

Finally, it should be pointed out that for values of frequency beyond 12000 Hz the sinusoidal signal started to look distorted on the screen of the smartphone used as oscilloscope.

### Final remarks

Two smartphones, one used as a signal generator, and another as an oscilloscope are good enough to carry out a quantitative study of the frequency curve in the RLC series circuit when it is connected to a sinusoidal electromotive force. The theoretical curve for the voltage at the resistor as a function of the frequency agrees very well with the experimental data. Similarly, the theoretical resonance frequency agrees well with the measured one. These results indicate the suitability of the smartphones for this type of AC current experiment. The low cost and portable experiment presented in this work can be easily implemented as a laboratory in secondary and first year physics courses contributing in this way to increase the motivation, independence, and academic performance of the students.